\documentclass[runningheads]{llncs}
\usepackage[T1]{fontenc}
\usepackage{array}
\usepackage{multicol}
\usepackage{multirow}
\usepackage{graphicx}
\usepackage{tabularray}
\usepackage{makecell} %for thicker hline table

\usepackage{graphicx}
\begin{document}

\title{How You Split Matters: Data Leakage and Subject Characteristics Studies in Longitudinal Brain MRI Analysis}
\titlerunning{How You Split Matters}

%\thanks{Supported by organization x.}}

\author{Dewinda J. Rumala\orcidID{0000-0002-1458-2238}} %/index{Rumala, Dewinda}
\institute{Institut Teknologi Sepuluh Nopember, Surabaya, Indonesia\\\email{dewinda.207022@mhs.its.ac.id}}
% \institute{Anonymous Organization\\ \email ********@***.***}
% \and
% \url{http://www.springer.com/gp/computer-science/lncs} 
% \and
% \footnote{corresponding author: dewinda.207022@mhs.its.ac.id}
\maketitle              % typeset the header of the contribution
\begin{abstract}
Deep learning models have revolutionized the field of medical image analysis, offering significant promise for improved diagnostics and patient care. However, their performance can be misleadingly optimistic due to a hidden pitfall called 'data leakage'. In this study, we investigate data leakage in 3D medical imaging, specifically using 3D Convolutional Neural Networks (CNNs) for brain MRI analysis. While 3D CNNs appear less prone to leakage than 2D counterparts, improper data splitting during cross-validation (CV) can still pose issues, especially with longitudinal imaging data containing repeated scans from the same subject. We explore the impact of different data splitting strategies on model performance for longitudinal brain MRI analysis and identify potential data leakage concerns. GradCAM visualization helps reveal shortcuts in CNN models caused by identity confounding, where the model learns to identify subjects along with diagnostic features. Our findings, consistent with prior research, underscore the importance of subject-wise splitting and evaluating our model further on hold-out data from different subjects to ensure the integrity and reliability of deep learning models in medical image analysis.

\keywords{Data Leakage \and Deep Learning \and MRI \and Alzheimer's Disease}
\end{abstract}
%
% -------------Input your contents here-------------------
\section{Introduction}

Medical image analysis has witnessed remarkable advancements with the integration of deep learning models, particularly Convolutional Neural Networks (CNNs), which have demonstrated great potential in enhancing diagnostics and patient care. Deep learning models have shown remarkable accuracy in classifying medical images, including brain MRI scans, and have become invaluable tools for medical professionals. However, the reliability of these models can be jeopardized by different biases  \cite{drukkerFairnessArtificialIntelligence2023}, especially a hidden challenge known as 'data leakage'.

Data leakage occurs when information from the test set unintentionally leaks into the training process \cite{kaufman2012leakage}, leading to over-optimistic performance during model evaluation \cite{bussolaAISlippingTiles2021}. This misleading optimism can result in a false perception of the model's efficacy, potentially compromising clinical decision-making and patient outcomes. Data leakage can arise from various sources, including improper data splitting strategies during cross-validation (CV) procedures \cite{yagisEffectDataLeakage2021,chaibubnetoDetectingImpactSubject2019}.

In this study, we focus on investigating data leakage in the context of 3D medical imaging, specifically using 3D CNNs for longitudinal brain MRI analysis. Longitudinal brain MRI data is particularly challenging due to repeated scans from the same subjects over time, introducing complex temporal dependencies that can exacerbate data leakage. While 3D CNNs have been considered less prone to data leakage than their 2D counterparts \cite{yagisEffectDataLeakage2021}, the impact of different data splitting strategies on model performance in this domain remains relatively unexplored.

Our primary objective is to examine the influence of data splitting strategies, specifically subject-wise, record-wise, and late-wise, on the performance of 3D CNN models in longitudinal brain MRI when employed for Alzheimer's Disaese (AD) classification. We aim to uncover potential issues related to data leakage and identity confounding as argued by \cite{chaibubnetoDetectingImpactSubject2019}, where the model learns to identify individual subjects rather than focusing solely on the diagnostic features of interest. To achieve this, we employ GradCAM visualization to gain insights into the attention patterns of the CNN models during classification.

This study contributes valuable insights into the impact of data leakage on deep learning models in the domain of medical image analysis, with a specific focus on longitudinal brain MRI data. Our findings are expected to shed light on the importance of proper data splitting strategies and the significance of subject-wise splitting in mitigating data leakage concerns. Ultimately, our research aims to enhance the integrity and reliability of deep learning models, promoting their seamless integration into clinical practice for longitudinal brain MRI classification.

\subsubsection{Related work.} 
When employing deep learning models for Alzheimer's disease (AD) diagnosis using MR images, many studies have traditionally relied on two-dimensional (2D)-based analysis using 2D CNNs. However, the slice-level analysis is prone to bias, including data leakage, as highlighted by Yagis et al. \cite{yagisEffectDataLeakage2021}. In response to these limitations, recent works have shifted towards using 3D CNNs for volume-based analysis, which has shown to be more superior compared to 2D-based analysis \cite{NarazaniIsAPet2022}. Several studies have demonstrated the advantages of 3D CNNs in capturing spatial information and improving the accuracy of medical image analysis \cite{goenkaAlzVNetVolumetricConvolutional2022a,ghazalAlzheimerRsquoDisease2018,zhang3DDenselyConnected2021,korolevResidualPlainConvolutional2017,NarazaniIsAPet2022}. 

Despite significant advancements and high accuracies achieved by deep learning models, many studies overlook the importance of proper data splitting procedures to avoid data leakage. Specifically, in longitudinal data, the occurrence of identity confounding should be avoided by using subject-wise split, as advocated by Neto et al. \cite{chaibubnetoDetectingImpactSubject2019} and Saeb et al. \cite{saebNeedApproximateUsecase2017}. Additionally, the commonly overlooked practice of late-wise split (where data splitting is performed after data transformation) has been identified as a potential source of data leakage, as suggested by Yagis et al. \cite{yagisEffectDataLeakage2021}. Thus, it should be avoided to ensure a reliable evaluation of deep learning models.
\section{Methods}

\subsection{Data Collection and Processing\label{sec:data}}
We obtained and used the preprocessed MRI data from the Alzheimer's Disease Neuroimaging Initiative (ADNI) \cite{jack2008AlzheimerDiseaseNeuroimaging} (Full details about the preprocessing step can be found at the ADNI website\footnote{http://adni.loni.usc.edu/methods/mri-analysis}), where we specifically selected T1-weighted and T2-weighted images acquired using a 3T scanner from the same demographic. To ensure data consistency, we applied strict inclusion and exclusion criteria and included only subjects with complete 3T MRI scans for both modalities. The collected dataset consists of longitudinal data from multiple visits for each subject, comprises 25 patients of AD, 41 patients of healthy controls (CN), and 45 patients with mild cognitive impairment (MCI). For each class, there are 150 scans in total, except for AD, where we collected only 50 scans.

The data was then further analyzed using the the Computational Anatomy Toolbox 12 (CAT12) \cite{gaserCATComputationalAnatomy2022} implemented in the Statistical Parametric Mapping 12 (SPM12) software. In CAT12, MRI images were processed using the usual VBM pipeline. As a result, all images have been rescaled to 113 $\times$ 137 $\times$ 113 and the intensity values are in range between 0 and 1. And to increase the number of data, we performed 3D augmentation techniques using the libraries provided by \cite{solovyev3DConvolutionalNeural2022}. The augmentation techniques performed include flipping and 5 degree rotation as previously practiced by \cite{goenkaAlzVNetVolumetricConvolutional2022a}. Thus, the final size of our data collection is 300 volume image scans for every class. Further detail about data statistics can be found in Table \ref{tab:datastat}.
% Toy example of data formation in this study can be found in Figure S1. 

\subsection{Training Setup}
We employed the widely recognized CNN architecture, DenseNet121, which was originally designed for 2D image analysis. To adapt it for 3D medical image analysis, we utilized the 3D architecture version of DenseNet121 as proposed by Soleovyel et al. \cite{solovyev3DConvolutionalNeural2022}. Based on their extensive investigation, 3DDenseNet121 demonstrated superior performance compared to other architectures when applied for 3D medical image analaysis task. In addition, we made a few modifications to the architecture to better suit our needs. 

Following the last 3D convolutional layer in the architecture, we incorporated a 3D global average pooling layer to reduce the spatial dimensions and obtain a fixed-length feature vector. This feature vector is then connected to fully connected layers with 2 units $\left \{128, 3 \right \}$, enabling us to perform three-way classification for the target classes of CN, MCI, and AD.

Between the fully connected layers, we incorporated rectified linear unit (ReLU) activation layers to introduce non-linearity and enhance the network's expressiveness. The final layer is a softmax classification layer, which provides probability scores for each class. We utilized the categorical cross-entropy loss function to train the model for the multi-class classification task. More information about the training setup is summarized in the supplemental Table S2.

\subsection{Evaluation Scheme\label{sec:scheme}}

Our main goal is to assess the impact of data leakage during CV by considering three different splitting strategies, each adjusted to our longitudinal data conditions, as detailed in Section \ref{sec:data}. To better illustrate the implementation, we provide a toy example of the three splitting strategies in Fig. \ref{fig:toy}. In this study, the dataset is then divided into train/validation/test sets with a rough ratio of 70/10/20\%, respectively. We conduct CV across 5-folds to thoroughly evaluate the performance and generalizability of our deep learning models under various splitting strategies.

To ensure an unbiased evaluation of our models and maintain a balanced representation of each class in each fold, we employ a stratified splitting strategy. This involves partitioning the images per class into k-folds, ensuring an equal distribution of images for each class across the folds. By doing so, we prevent any class-specific biases during training and evaluation, thus providing a fair assessment of our deep learning model's performance.

These three evaluation schemes allow us to comprehensively investigate and understand the potential impact of data leakage on the model's performance under different data splitting conditions. This provides a holistic assessment of the deep learning model's reliability and generalizability

\subsubsection{Subject-Wise Split.}
% This strategy will be regarded as the proper way to split longitudinal data as suggested by Neto, et al and Yagis, et al.
In this scheme, all image scans of each subject are assigned as a group in a fold, regardless of their visit times before any data transformation is applied (early splitting). This approach ensures that the scans from the same subject are not spread across different folds but rather kept together, allowing the model to be evaluated on unseen subjects during each fold. This helps to avoid any potential data leakage and identity confounding between subjects as suggested by Neto, et al. \cite{netoAnalysisPersonalizedMedication2017}.
% Considering that all data available for each class are split based on the subject regardless of the visit time and augmented version. This practice is to ensure that any data from the same subject would not be seen in different sets (training, validation, and test sets). Hence, when applying this splitting strategy, the sets of data for each fold of each class can be written such as in equation 1.

\subsubsection{Record-Wise Split.}
In this scheme, image scans from all subjects are grouped together into a fold based on the records or visits time in an early split manner. With this strategy, data from different visits of the same subject may appear in different folds, allowing the model to be trained and evaluated on the same subjects during each fold. While some studies \cite{littleUsingUnderstandingCrossvalidation2017} believe that this strategy might be conditionally appropriate and can lead to better results compared to subject-wise fashion data split, it may not be the best approach and can lead to overconfident performance due to data leakage, specifically caused by the occurrence of identity confounding as argued by Neto et al. \cite{netoAnalysisPersonalizedMedication2017}.

\subsubsection{Late Split.} 
\begin{figure}[!ht]
\renewcommand{\figurename}{Fig.}
\renewcommand{\thefigure}{S1}
\includegraphics[width=\textwidth]{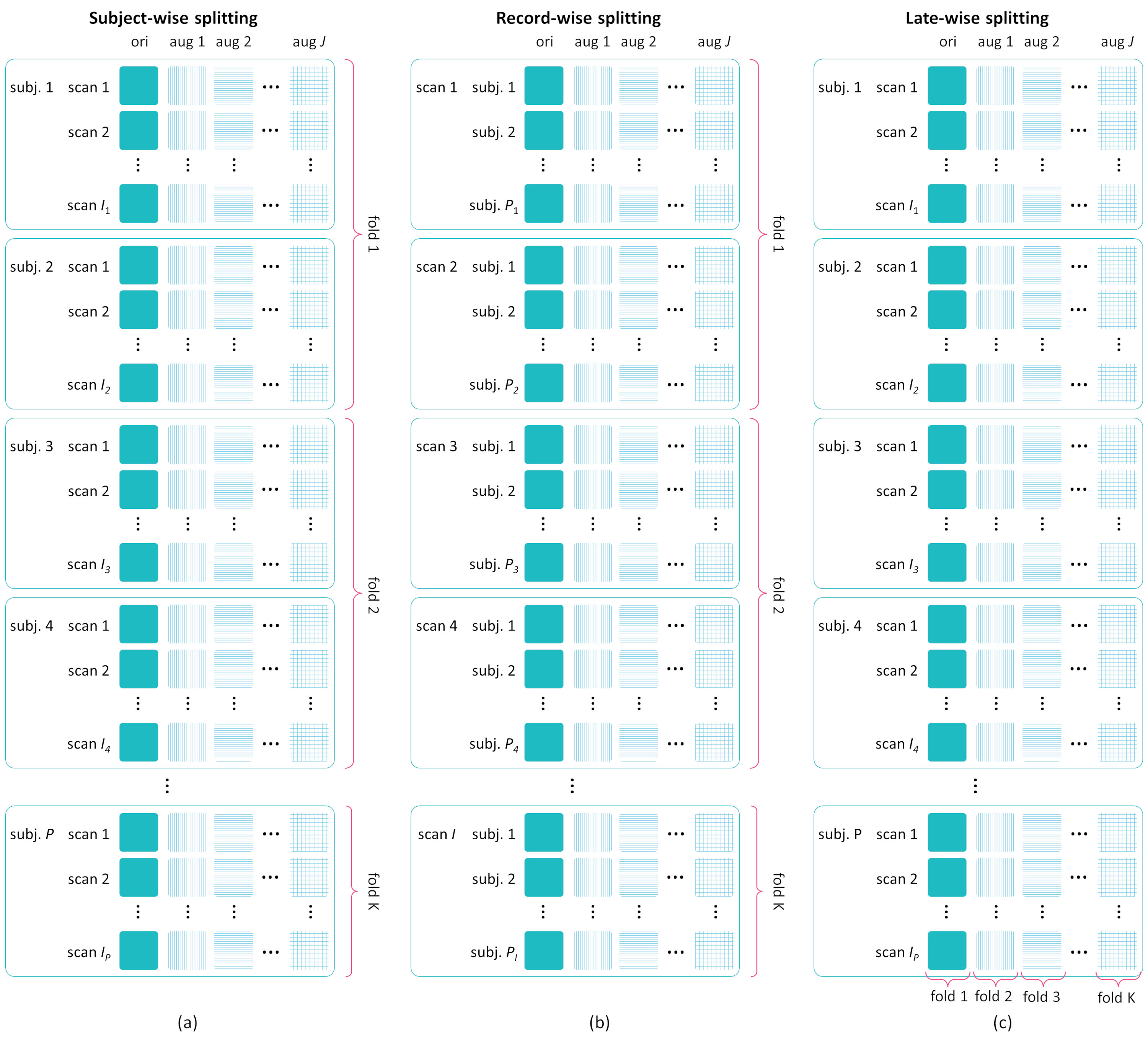}
\caption{A toy example of different data split strategies for longitudinal brain MRI. (a) Subject-wise splitting groups all image scans based on the subjects into k-folds. (b) Record-wise splitting groups image scans based on different visit times into k-folds. (c) Late-wise splitting groups image scans based on transformation technique into k-folds.} \label{fig:toy}
\end{figure}

Data splitting is performed after the augmentation process in this scheme, by using a sequential numbering approach (more information described in Fig. \ref{fig:toy}). This split allows the augmented image scans of the same subjects from a certain visit to appear in different folds. According to previous studies by Varoquaux et al. \cite{varoquauxMachineLearningMedical2022} and Yagis et al. \cite{yagisEffectDataLeakage2021}, this splitting strategy has been shown to cause data leakage and should be avoided.

\section{Result}
% In this section, we present the comprehensive evaluation results of our proposed method using three different data splitting strategies: subject-wise, record-wise, and late-wise to classify images into CN, MCI, and AD on longitudinal brain MRI data. We begin by presenting the performance of our model on the 5-folds cross-validation data, followed by additional evaluation on an expanded dataset with more subjects. Finally, we provide visual insights into the model's behavior using GradCAM visualization. The primary focus is to how splitting strategies can affect the model's ability to classify whole brain MR images accurately.

\subsubsection{Evaluation Results on 5-Fold Data.}

We first investigated the impact of different data splitting strategies towards model performance on the 5-fold data. Table \ref{tab:interntest} reports the results for the experiments described in section \ref{sec:scheme} on different MRI sequences of T1-weighted and T2-weighted MRI. As seen in the table, the performance of the models varied across the different splitting strategies.

Upon comparing the performance across various data splitting strategies, a significant difference was observed (P=0.0389), with the record-wise strategy achieving the highest mean accuracy across all MRI sequences. Conversely, the subject-wise strategy obtained the lowest mean accuracy when evaluated on both sequences.

Furthermore, our investigation revealed no statistically significant difference between the T1-weighted and T2-weighted MRI sequences in terms of model performance (P=0.7921). This finding suggests that both sequences are equally suitable for the three-way classification task of CN, MCI, and AD. Consequently, the choice of MRI sequence does not significantly influence the overall classification performance.

\begin{table}[!ht]
\centering
\caption{Training and testing on longitudinal T1-weigted and T2-weighted MRI. Values are presented as mean and standard deviation across folds, expressed in percentages.}\label{tab:interntest}
\begin{tabular}{cccccc}
\Xhline{2\arrayrulewidth}
MRI Sequence & Scheme & Acc & Prec& Rec & F1-score\\ \hline
\multirow{3}{*}{T1-weighted}& Subject-wise & $67.11 \pm 6.11$ & $69.38 \pm 6.02$ & $67.11 \pm 6.12$  & $68.28 \pm 5.63$\\
&Record-wise & $97.33 \pm 1.86$ & $97.54 \pm1.66$ & $97.33 \pm 1.86$  & $97.34 \pm 1.85$\\
&Late-wise & $81.33 \pm 12.37$ & $89.45 \pm 8.31 $ & $79.31 \pm 13.29$  & $89.44 \pm 77.64$\\ \hline
\multirow{3}{*}{T2-weighted}& Subject-wise & $61.56 \pm 5.23$ & $63.81 \pm 9.56$ & $62.56 \pm 5.59$  & $61.55 \pm 5.77$\\
&Record-wise & $93.33 \pm 1.55$ & $95.53 \pm 1,45$ & $95.33 \pm 1.55$  & $95.42 \pm 1.77$\\
&Late-wise & $88.89 \pm 6.30$ & $89.32 \pm 6.34$ & $88.89 \pm 6.30$  & $88.75 \pm 6.37$\\ 
\Xhline{2\arrayrulewidth}

\end{tabular}
\end{table}

\subsubsection{Further Evaluation on More Subjects.}
We evaluated the robustness of 3D CNN models trained with different splitting strategies using a separate hold-out dataset of longitudinal brain MRI. The hold-out data consists of image scans from subjects with different visits, including 8 CN, 11 MCI, and 7 AD subjects, totaling 30 image scans per class (see Table \ref{tab:datastat} for details). Due to data availability constraints, the evaluation focused on T1-weighted MRI images, but future research could benefit from including T2-weighted MRI images for a more comprehensive assessment.

From the previous training and evaluation of the model on 5-fold data, we obtained five different models for each data split strategy that were directly evaluated on the hold-out data. The results of this robustness evaluation are presented in Table \ref{tab:robust}. From this investigation, we discovered notable discrepancies between the 5-fold and hold-out data evaluation for all splitting strategies. 

For instance, the record-wise strategy, which demonstrated an impressive mean accuracy of 97.33\% during cross-validation, experienced a substantial drop to 38.71\% when tested on the hold-out data. Similarly, the late-wise strategy, initially yielding a mean accuracy of 81.33\% during cross-validation, exhibited a significant decline to 40.43\% in the robustness evaluation. Conversely, subject-wise splitting displayed a least drastic drop in accuracy compared to the other splitting strategies, from 67.12\% to 42.15\%. 

We observed no statistically significant difference between data split strategies on hold-out data ($P=0.8235$). However, it is worth noting that subject-wise splitting strategy achieved the highest mean accuracy in this evaluation, while record-wise splitting obtained the lowest mean accuracy. These results contrast with the model's performance on 5-fold data.

\begin{table}[!ht]
\centering
\caption{Robustness evaluation of trained models on hold-out data of T1-weighted MRI. Values are presented as mean and standard deviation obtained from models trained previously on different folds, expressed in percentages.}
\label{tab:robust}
\begin{tabular}{ccccc}
\Xhline{2\arrayrulewidth}
Splitting Scheme & Acc & Prec& Rec & F1-score\\ \hline
Subject-wise & $42.15 \pm 5.45$ & $38.71 \pm 7.54$ & $42.12 \pm 5.50$  & $38.57 \pm 4.99$\\
Record-wise & $38.71 \pm 7.75$ & $37.48 \pm 9.20$ & $38.63 \pm 7.72$  & $35.68 \pm 7.37$\\
Late-wise & $40.43 \pm 8.95$ & $37,62 \pm 13.31 $ & $40.43 \pm 8.95$  & $39.92 \pm 4.80$\\
\Xhline{2\arrayrulewidth}
\end{tabular}
\end{table}

\subsubsection{Grad-CAM Visualization.}

\begin{figure}[!ht]
\includegraphics[width=0.6\textwidth]{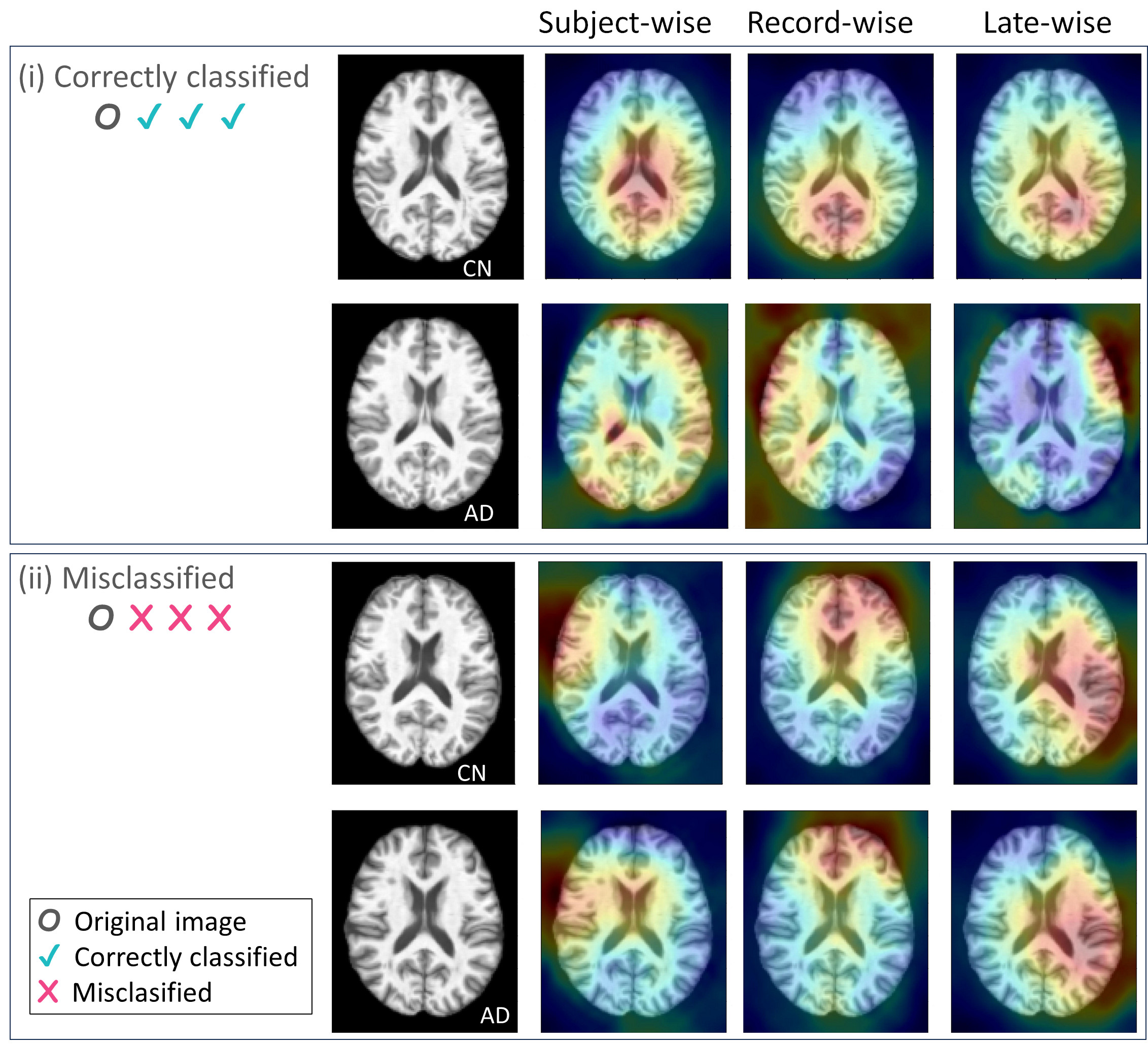}
\centering
\caption{GradCAM visualization examples for Normal Control (CN) and Alzheimer's Disease (AD) classes under different data split schemes on axial slices of T1-weighted MRI. Each pair of rows depicts (i) correctly classified images, and (ii) misclassified images across all splitting strategies} \label{fig:gradcam}
\end{figure}

We present in Fig. \ref{fig:gradcam} examples of the gradient class activation maps (Grad-CAM) for the CN and AD classes extracted and fused from all layers of the 3D CNN models. GradCAM visualization was performed on the hold-out data to gain insights into the attention patterns during classification for the 3D CNN models trained using different data split strategies.

The first row in Fig. \ref{fig:gradcam} displays GradCAM examples of correctly classified CN and AD images from different splitting strategies. We observe that the patterns for achieving correct predictions on CN are relatively similar across the splitting strategies, where some activations are shown around middle area of the brain. However, there are notable differences between the attention patterns for correctly classified AD images under each splitting strategy.

The Grad-CAM results from the subject-wise splitting strategy show activations around the peripheral and middle regions of the brain, contributing to the correct identification of AD. However, when employing the record-wise splitting strategy, the activations become more scattered compared to using subject-wise splitting. While some activated areas remain shown in the left hemisphere, there are more irrelevant activated regions outside of the brain. 

The same pattern is observed for the model trained using the late-wise splitting strategy, with even worse results. In this case, the activated areas are heavily concentrated in the corner of the right brain. The results from the record-wise and late-wise strategies may indicate that the models have learned hidden patterns that may not be easily understood by radiologists, rather than focusing on the expected brain patterns for detecting AD.

Moving to the second row, which showcases misclassified images, we can observe similarities in the attention patterns across the different data split strategies. Notably, the heatmaps tend to drift off to the background and do not heavily highlight regions in the brain, which suggests challenges in accurately classifying these images.

% In the case of subject-wise classification, the GradCAM heatmaps revealed clear and interpretable patterns. The model focused its attention on specific brain regions known to be associated with the target classes. This behavior indicated that the subject-wise splitting strategy provided a more consistent and meaningful representation of class-specific features, leading to more reliable interpretations.

% Conversely, when using record-wise and late-wise splitting strategies, the GradCAM heatmaps showed patterns that were less interpretable and often extended beyond the brain's relevant regions. This suggests that these strategies might introduce shortcuts and capture non-informative features, leading to potential data leakage and overconfident predictions.

% Notably, the record-wise and late-wise splitting strategies exhibited increased attention to irrelevant areas, indicating the influence of identity confounding and potential biases introduced during data splitting.

% Overall, the GradCAM visualization results emphasize the importance of subject-wise splitting in maintaining a reliable and interpretable model, while highlighting the risk of data leakage and shortcut learning associated with record-wise and late-wise splitting strategies.

\section{Discussion and Conclusion}

In this study, we conducted a comprehensive evaluation of the effect of data leakage on a 3D CNN model for classifying longitudinal brain MRI data into CN, MCI, and AD using three data splitting strategies: subject-wise, record-wise, and late-wise. Our investigation began with 5-fold data, revealing significant differences in accuracy across the splitting strategies.

The record-wise strategy performed impressively during cross-validation, while subject-wise showed the lowest accuracy. Surprisingly, when tested on hold-out data, all strategies experienced significant drops in accuracy. The record-wise strategy, initially the best, suffered the worst decline. Therefore, to gain further insights into the model's behavior, we performed GradCAM visualization of the model evaluated on the hold-out data. 

The activation heatmaps for correctly classified images showed that there might be shortcut learning in record-wise and late-wise splitting to correctly identify AD. This means that the models learned unintended patterns during cross-validation, which do not generalize well to new, unseen data \cite{geirhosShortcutLearningDeep2020,jimenez2022detecting}. This might explain why there is significant drop in record-wise and late-wise strategies. As argued by Neto, et al. \cite{chaibubnetoDetectingImpactSubject2019}, such record-wise strategy is prone for identity confounding, where the model try to learn identifying the subject rather than target class. Meanwhile, late-wise splitting was also found to cause data leakage \cite{yagisEffectDataLeakage2021}. Besides, it may might also experience identity confounding. 

Meanwhile, in subject-wise splitting, the model shows more interpretable heatmaps, indicating that it is focusing on more relevant regions for AD diagnosis and not relying on shortcuts or irrelevant information. However, there is a decrease in performance when tested on hold-out data compared to cross-validation data, suggesting that the model might still be facing some challenges, which according to Little, et al. \cite{littleUsingUnderstandingCrossvalidation2017}, subject-wise splitting strategies can lead to model under-fitting and larger classification errors. The occurence of under-fitting for this strategy makes sense, as the model showed unnecessary activations on irrelevant areas of misclassified images, which suggests that the model may not be learning all the relevant features needed for accurate diagnosis. 

However, the less significant drop in performance compared to record-wise and late-wise splitting strategies indicates that subject-wise split might be more robust and less prone to data leakage. And as suggested by Neto, et al. \cite{chaibubnetoDetectingImpactSubject2019}, overcoming model under-fitting when applying subject-wise split can be addressed by having more subjects in the dataset. With a larger and more diverse dataset, the model can learn from a broader range of samples, enabling it to capture more complex patterns and improve its performance and generalization on both cross-validation and hold-out data.

We acknowledge certain limitations in this study, particularly the issue of under-fitting due to the limited number of subjects in the subject-wise approach. Additionally, we recognize the gender imbalance, as the study includes a higher number of female participants compared to male participants. For a less biased analysis, future studies should aim to include a larger and more balanced representation of participants in terms of gender and other demographic factors \cite{riccilaraAddressingFairnessArtificial2022}. Moreover, exploring the effects of different splitting strategies on various demographic groups would be essential for ensuring a stronger fairness in the deep learning model's performance.

In summary, this study highlights the significant impact of data leakage on model performance, leading to over-optimistic results. Our findings align with previous research advocating for the use of subject-wise splitting approach \cite{netoAnalysisPersonalizedMedication2017,saebNeedApproximateUsecase2017,chaibubnetoDetectingImpactSubject2019} and early-split \cite{yagisEffectDataLeakage2021} to avoid data leakage and identity confounding. Moreover, it is advisable to incorporate hold-out data from different subjects whenever possible, in accordance with the recommendations of Varoquaux et al. \cite{varoquauxMachineLearningMedical2022}. Implementing these suggestions can improve deep learning model robustness and reliability in medical image analysis, particularly for longitudinal brain MRI classification.

% \subsubsection{Acknowledgements} Please place your acknowledgments at the end of the paper, preceded by an unnumbered run-in heading (i.e. 3rd-level heading).
\subsubsection{Acknowledgements} This research was funded by the Ministry of Education and Research Technology, Indonesia. Special thanks to Dr. I Ketut Eddy Purnama, the author's PhD supervisor, for securing the research funding. Further appreciation to Prof. Tae-Seong Kim, whose insightful perspectives inspired the development of this paper, and to the Bio Imaging Laboratory at Kyung Hee University, South Korea, where the data collection for this study was conducted.


\bibliographystyle{splncs04}
\bibliography{ref}

%----supplementary method here-------
\clearpage
\section*{Supplementary Material}

\begin{table}
\renewcommand{\tablename}{Table}
\renewcommand{\thetable}{S1}
\centering
\caption{Dataset Statistics. Ratio of female/male is based on number of images rather than number of subjects.}
\label{tab:datastat}
\begin{tabular}{ c c c c c c c}
\Xhline{3\arrayrulewidth}
\multirow{2}{*}{Collection} & \multirow{2}{*}{Groups} & \multirow{2}{*}{No of Subj.} &\multirow{2}{*}{Female/Male} & \multirow{2}{*}{Age} & \multicolumn{2}{c}{No of Scans} \\ \cline{6-7}
& & & & & Before Aug   & After Aug \\ \hline
\multirow{3}{*}{5-fold data}  
& CN & 41 & 85/65  & $76.68 \pm 4.15$& 150 & 300 \\
& MCI & 45 & 50/100 & $74.19 \pm 8.57$ & 150 & 300\\
& AD & 25 & 30/20  & $74.22 \pm 8.90$ & 50 & 300\\ \hline

\multirow{3}{*}{Hold-out data}  
& CN & 8 & 17/13  & $74.81 \pm 3.13$  & 30 & - \\
& MCI & 11 & 8/22 & $75.42 \pm 7.26$ & 30 & -\\
& AD & 7 & 16/1 & $78.26 \pm 6.52$  & 30 & -\\

\Xhline{2\arrayrulewidth}
\end{tabular}
\end{table}

\begin{table}
\renewcommand{\tablename}{Table}
\renewcommand{\thetable}{S2}
\centering
\caption{Overview of the parameters used across all experiments: the learning rate was reduced using the ReduceLROnPlateau scheduler from TensorFlow by a factor of 0.1 when validation loss did not decrease after 10 epochs. Adam was used as the optimizer.}
\label{tab:hyper}
\begin{tabular}{ccccc}
\Xhline{2\arrayrulewidth}
Parameter & Value\\ \hline
Learning rate & 0.0001 \\
Epsilon & 0.0001\\
Beta 1 & 0.9\\
Beta 2 & 0.99\\
Epoch & 100 \\
Batch size & 24 \\

\Xhline{2\arrayrulewidth}
\end{tabular}
\end{table}

\end{document}